\documentclass[prl,twocolumn,showpacs,graphicx,floatfix]{revtex4}
\usepackage{epsf}

\begin{document}

\title{Metal-Insulator-Like Behavior in Semimetal{\bf lic} Bismuth and Graphite}
\author{Xu Du, Shan-Wen Tsai\cite{tsai}, Dmitrii L. Maslov, and Arthur F. Hebard}
\affiliation{Department of Physics, University of Florida, P. O. Box 118440, Gainesville,
FL 32611-8440}

\begin{abstract}
When high quality bismuth or graphite crystals are placed in a magnetic field directed along
the c-axis (trigonal axis for bismuth) and the temperature is lowered, the resistance
increases as it does in an insulator but then saturates.
We show that the combination of unusual features specific to semimetals, i.e., low
carrier density, small effective mass, high purity, and an equal number of electrons and holes
(compensation), gives rise to a unique ordering and spacing of three characteristic energy
scales, which not only is
specific to semimetals but which concomitantly provides a wide window for the observation of
apparent field induced metal-insulator behavior. Using magnetotransport and Hall
measurements, the details of this unusual behavior
are captured with a conventional multi-band model, thus confirming the occupation by
semimetals of a unique niche between conventional metals and semiconductors.
\end{abstract}

\date{\today}
\maketitle
Elemental semimetals, such as bismuth and graphite, are intriguing
materials to study because of their high magnetoresistance, low
carrier density $n$, and high purity. Due to small values of $n$,
magnetic fields on the order of 10~T are sufficient to drive these
semimetals into the ultraquantum regime, where only the lowest
Landau level remains occupied. In addition, light cyclotron
masses, $m^{\ast}$, for any field orientation in Bi and along the
c-axis in graphite, result in higher cyclotron frequencies,
$\omega_c=eB/m^{\ast}$, which ensure that quantum
magneto-oscillations can be observed at moderate temperatures.
High purity
facilitates the observation of well-resolved oscillation patterns.
These features have made bismuth and graphite perhaps the two most
popular materials for studies of quantum magnetic-field effects
\cite{edelman,brandt}.

Recently, interest in magnetotransport in graphite has been
renewed due to observations (similar to those shown in Fig.~1) of
an effect that looks like a magnetic-field-induced metal-insulator
transition \cite{kop1,kop2}. The metallic $T$-dependence of the
in-plane resistivity in zero field turns into an insulating-like
one when a magnetic field on the order of 10~mT is applied normal
to the basal (ab) plane. Increasing the field to about 1~T
produces a re-entrance of the metallic behavior \cite{kop3}. It
has been proposed that the low-field effect is due to a
magnetic-field-induced excitonic insulator transition of Dirac
fermions \cite{kop3,khv}, whereas the high-field one is a
manifestation of field-induced superconductivity \cite{kop3}. It
has been also suggested \cite{kop3} that the apparent
metal-insulator transition in graphite is similar to that in 2D
heterostructures \cite{kravchenko} (although the latter is driven
by a field parallel to the conducting plane).

Similar metal-insulating like behavior is also observed in
99.9995\% pure bulk bismuth
crystals as shown in Fig.~2, where the resistivity is plotted as a
function of temperature in successively higher magnetic fields.
The crossover from ``metallic'' to ``insulating'' behavior has the
same qualitative behavior in both semimetals. On lowering the
temperature the resistance increases, as it does in a insulator,
but then saturates towards field-dependent constant values at the
lowest temperatures. These similarities invite an interpretation
that ascribes this interesting behavior to properties shared by
both graphite and bismuth, namely low carrier density, high
purity, and an equal number of electrons and holes (compensation),
rather than to specific properties of graphite: almost 2D
nature of transport and a Dirac-like spectrum, as suggested
in Refs.\cite{kop1,kop2,khv,kop3}.

In this letter we demonstrate this connection by examining the
magnitude and ordering of three characteristic energy scales:
namely, the 
width of the energy levels $\hbar /\tau$ where $\tau$ is the
electron-phonon scattering time, the cyclotron energy $ \hbar
\omega _{c}$,
and the thermal energy $k_B T$. We provide theoretical
justification for, and experimental confirmation of, the existence
of a wide interval of temperatures and magnetic fields, defined by
the condition,
\begin{equation}
\hbar /\tau \lesssim \hbar \omega _{c}\lesssim k_{B}T.  \label{a1}
\end{equation}
In this interval, (a) the magnetoresistance is large, (b) the
scattering rate is linear in $T$, and (c) Shubnikov de Haas (SdH)
oscillations are not resolved due to the thermal smearing of
Landau levels. We argue that the unusual behavior of bismuth and
graphite is due to the existence of a region, specified by the
inequalities of Eq. (\ref{a1}), and also due to compensation between
electron and hole charge carriers. 
Our experimental confirmation is centered on a
detailed study of graphite in which we use the conventional theory
of multi-band magnetotransport \cite{ashcroft} to extract the
field-independent carrier density, $n (T)$, and scattering time,
$\tau (T)$, from simultaneous fitting of the temperature and
field-dependent longitudinal resistivity $\rho_{xx} (T,B)$
(magnetoresistance) and transverse resistivity $\rho_{xy} (T,B)$
(Hall effect). We then show from this analysis that the inequality
(\ref{a1}), which is unique to semimetals, is satisfied over a
broad temperature and field range.

To illustrate the uniqueness of low-carrier-density semimetals, we
compare them with conventional, high-density, uncompensated
metals. To begin with, if the Fermi surface is isotropic, a metal
exhibits no magnetoresistance because the Lorentz force does not
have a component along the electric current \cite{ashcroft}. In
anisotropic metals the magnetoresistance is finite and
proportional to $(\omega _{c}\tau )^{2}$ in weak magnetic fields,
i.e., for $\omega _{c}\tau \ll 1$. In stronger fields ($\omega
_{c}\tau \gg 1$), classical magnetoresistance saturates, if the
Fermi surface is closed \cite{abrikosov}.
In contrast, transverse magnetoresistance of a semimetal grows as
$B^{2}$ both in the weak- and strong-field regimes
\cite{abrikosov}.

The magnetoresistance $[\rho(B)-\rho(0)]/\rho(0)$ is much larger in
semimetals than in conventional metals. In addition to the saturation effect,
described above, another important factor that limits the magnetoresistance
in conventional metals is the higher scattering rates and thus smaller values of
the $\omega_c\tau$ product. The impurity scattering rate in semimetals is
smaller than in conventional metals simply because semimetals are typically
much cleaner materials. The lower carrier density of semimetals also
reduces the rates of electron-phonon
scattering in semimetals  compared to that of conventional metals.
For temperatures
above the transport Debye temperature, which separates the regions of the
$T$- and $T^{5}$-laws in the resistivity, $\Theta _{D}^{\rho }=2\hbar
k_{F}s/k_{B}$, where $k_{F}$ is the Fermi wavevector and $s$ is  the speed
of sound (both properly averaged over the Fermi surface), one can estimate
the electron-phonon scattering rate as $\tau ^{-1}\simeq \left(
k_{F}a_{0}\right) \left( m^{\ast }/m_{0}\right) k_{B}T/\hbar $,
where $a_{0}$ is the atomic lattice constant, and $m^{\ast }$ and $m_{0}$ are
respectively the effective and bare electron masses \cite{levinson}. In a conventional
metal, $k_{F}a_{0}\simeq 1$ and $m^{\ast }\simeq m_{0}$. In this case, $%
\Theta _{D}^{\rho }$ is of order of the thermodynamic Debye temperature ~$%
\hbar s/k_{B}a_{0}\simeq $ few 100 K and $\tau ^{-1} \simeq  k_{B}T/\hbar$.
In low-carrier-density materials
($k_{F}a_{0}\ll 1$), which typically also have light carriers ($m^{\ast
}\ll m_{0}$), $\Theta _{D}^{\rho }$ is much smaller
and thus $\tau ^{-1}\ll k_{B}T/\hbar $.
Therefore, in a low-carrier-density semimetal there exists a wide interval
of temperatures and magnetic fields, defined by the inequalities (\ref{a1}).
In contrast, there is no wide interval between $\hbar /\tau $ and
$k_{B}T$ in a conventional metal \cite{comment}.

An additional feature, crucial for interpreting the experimental
data, is that the Fermi energies of graphite ($E_{F}$ = 22 meV)
\cite {mcclure} and bismuth ($E_{F}$ = 30 meV [holes])\cite{smith}
are relatively low and the temperature dependence of the
resistivity comes from two temperature-dependent quantities:
$n(T)$ and $\tau (T)$. Purity of materials ensures that
electron-phonon scattering is a dominant mechanism for resistance
over a wide temperature range.

\begin{figure}[tbp]
\begin{center}
\epsfxsize=1 \columnwidth
\epsffile{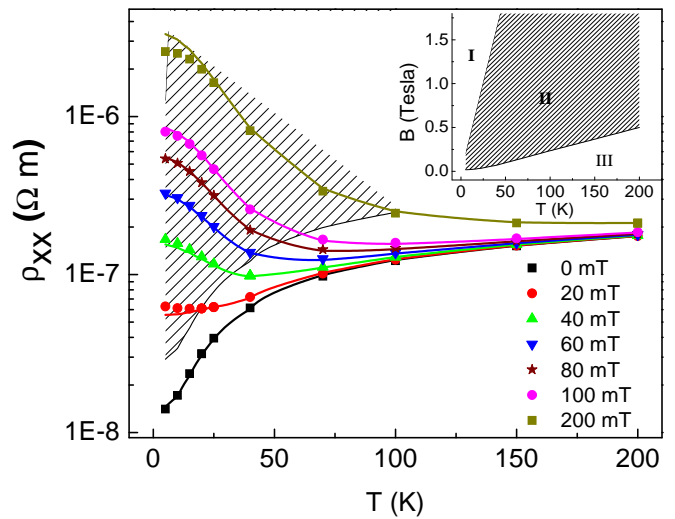}
\end{center}
\caption{ Temperature dependence of the ab-plane resistivity
$\rho_{xx}$ for a graphite crystal
at the c-axis magnetic fields indicated in the legend.
The solid lines are the fits to the data using the six parameters
derived from the three-band analysis described in the text. The
shadowed region in the inset and its mapping onto the data in
the main panel indicates the interval defined by Eq.(\ref{a1}).
}
\label{fig1}
\end{figure}

Standard 4-probe measurements were carried out on single-crystal highly
oriented pyrolytic graphite (HOPG) sample with a 2$^{\circ }$ mosaic spread,
as determined by X-rays.
The resistivity was measured using an ac (17 Hz) resistance bridge
over the temperature range 5K-350K. In all the measurements, the
magnetic fields were applied perpendicular to the sample basal
planes. Both $\rho _{xx}$ and $\rho _{xy}$ (Fig.~3) were measured
in magnetic fields up to 1~T, although the analysis (solid lines)
was limited to $B \leq 200$~mT. A small field-symmetric component
due to misaligned electrodes was subtracted from the $\rho
_{xy}\left( B\right) $ data.

\begin{figure}[tbp]
\begin{center}
\epsfxsize=1 \columnwidth
\epsffile{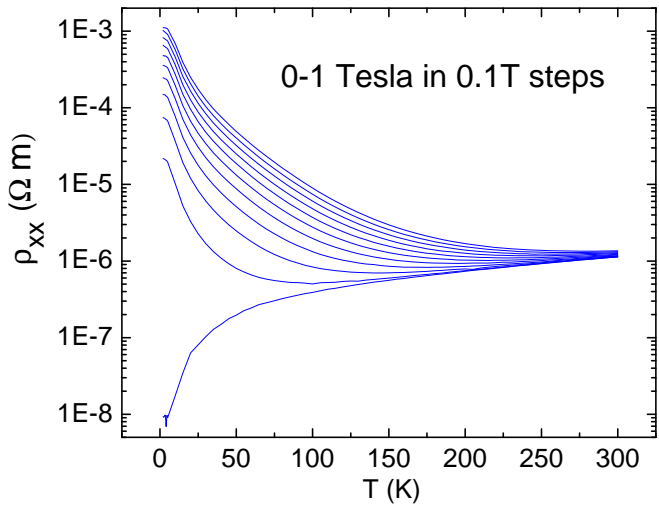}
\end{center}
\caption{
Temperature dependence of the longitudinal resistivity
for a bismuth crystal at magnetic fields ranging from
0 to 1~T in 0.1~T steps from top to bottom. The
magnetic field and the current are along the trigonal and binary
axes, correspondingly. In zero field, the resistance is
approximately linear in temperature. 
}
\label{fig1}
\end{figure}

\begin{figure}[tbp]
\begin{center}
\epsfxsize=1 \columnwidth
\epsffile{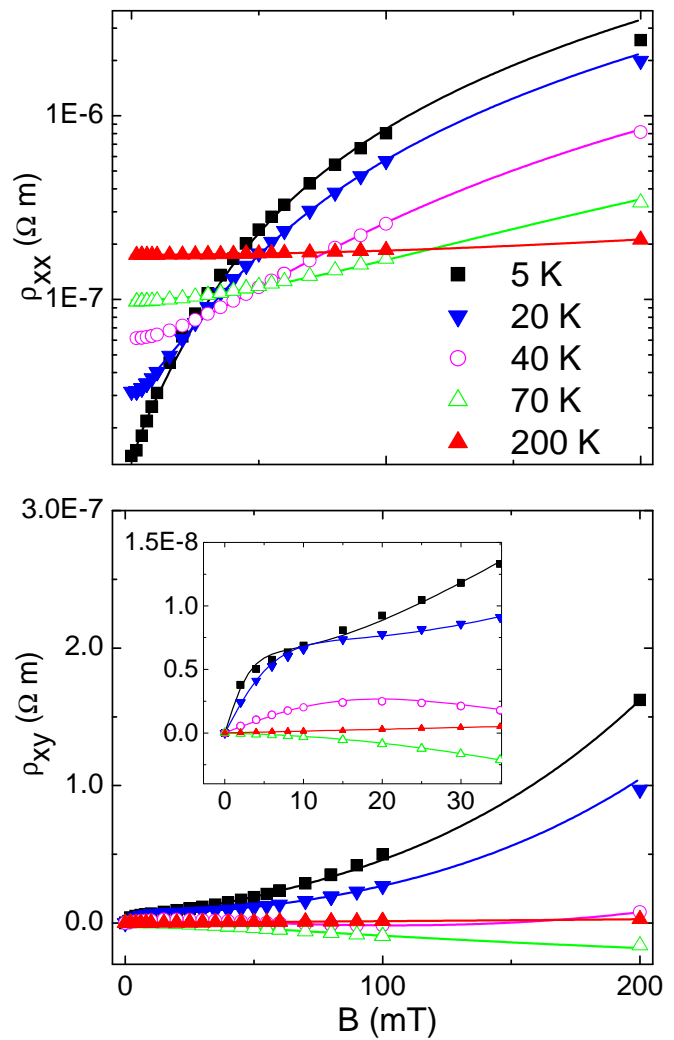}
\end{center}
\caption{
Longitudinal resistivity $\rho _{xx}$ (top panel)
and transverse resistivity  $\rho _{xy}$  (bottom panel)
versus applied field $B$ for graphite at the temperatures indicated in the
legend. The solid lines are determined by a fitting procedure that
simultaneously includes both sets of data into a three-band model
described in the text. The inset in the bottom panel, with the
same units on each axis, magnifies the low-field region, where the
contribution from the third band with a lower carrier density
makes a major contribution
that cannot be fit by solely using the two majority bands.
} \label{fig1}
\end{figure}

We used a standard multi-band model \cite{ashcroft} to fit the data. Each band has two
parameters: resistivity $\rho _{i}$ and Hall coefficient $R_{i}=1/q_{i}n_{i}$,
where $q_{i}=\pm e$ is the charge of the carrier. In agreement with earlier
studies, we fix the number of bands to three \cite{brandt}.  Two of these
are the majority electron and hole bands, and the third is the minority hole
band \cite{brandt}. Although the third band is not essential
for a qualitative understanding of the data, it is necessary for explaining
the low-field fine features in $\rho _{xy}$
shown in the inset of Fig~3.
The minority band makes a negligible contribution at higher
fields due to its low carrier concentration.

We fit $\rho _{xx}$ and $\rho _{xy}$
simultaneously by adjusting the six parameters
independently, until differences between the fitting curves and the
experimental data are minimized. Because the majority carriers in graphite
derive from Fermi surfaces that have six-fold rotational symmetry about the
c-axis, we only need to use the 2x2 magneto-conductivity tensor
$\hat{ \sigma}^{i}$ with elements $\sigma _{xx}^{i}=\sigma _{yy}^{i}=\rho _{i}/%
\left[ \rho _{i}^{2}+\left( R_{i}B\right) ^{2}\right] $
and $\sigma _{xy}^{i}=-\sigma _{yx}^{i}=-R_{i}B/\left[ \rho _{i}^{2}+\left(
R_{i}B\right) ^{2}\right] ,$
where $\rho _{i}=m_{i}^{\ast }/n_{i}e^{2}\tau _{i}$.
The total conductivity, $\hat{\sigma}=\sum_{i=1}^{3}\hat{\sigma}^{i}$,
is simply a sum of the contributions from all the bands and the total
resistance is $\hat{\rho}=\hat{\sigma}^{-1}$.

Qualitatively, the unusual temperature dependence displayed in
Fig. 1 can be understood for a simple two-band case
where $\rho _{xx}$ reduces to
\begin{equation}
\rho _{xx}=\frac{\rho
_{1}\rho _{2} \left( \rho _{1} + \rho _{2} \right)
+ \left( \rho _{1}R_{2}^{2}+\rho _{2}R_{1}^{2} \right) B^{2}}{\left( \rho _{1}+\rho
_{2}\right) ^{2}+\left( R_{1}+R_{2}\right) ^{2}B^{2}}.  \label{a2}
\end{equation}
Assuming that $\rho _{1,2}\propto T^{a}$ with $a>0$, we find that
for perfect compensation, $R_{1}=-R_{2}=|R|$, Eq.~(\ref{a2}) can
be decomposed into two contributions: a field-independent term
$\propto T^{a}$ and a field-dependent term $\propto
R^{2}(T)B^2/T^{a}$. At high $T$, the first term dominates and
metallic behavior ensues. At low $T$, $R(T) \propto 1/n(T)$
saturates and the second term dominates, giving ``insulating"
behavior.

\begin{figure}[tbp]
\begin{center}
\epsfxsize=1 \columnwidth
\epsffile{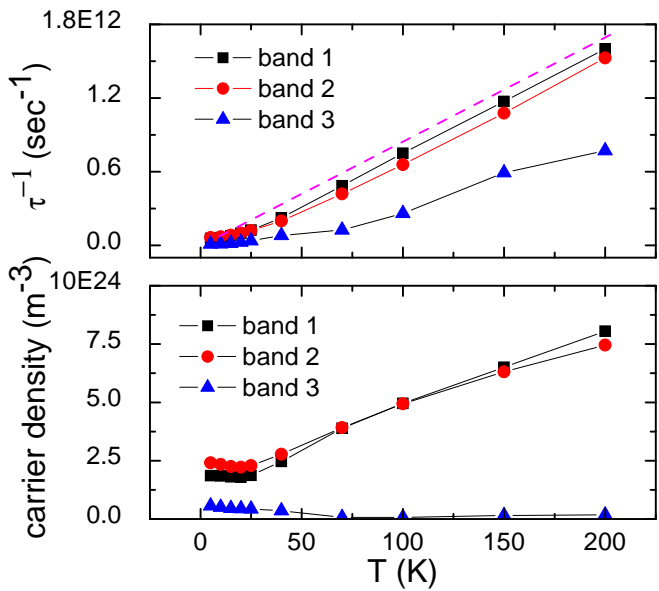}
\end{center}
\caption{
Temperature dependence of the fitting parameters (graphite) for the bands 
indicated in
the legends of each panel. The near equality of the carrier densities of the
two majority bands (lower panel) indicates good compensation at low fields
($B < 200$ mT) and the linear dependence of the scattering rate on $T$
(upper panel, dashed line) with a slope of
0.065(3) in units of $k_B/\hbar$ is consistent
(see text) with electron-phonon scattering.
}
\label{fig1}
\end{figure}

The actual situation is somewhat more complicated
due to the $T$-dependence of the carrier concentration,
the presence of the third band, and imperfect compensation between the
majority bands. Results for the temperature-dependent fitting parameters are
shown in Fig.~4 where band 1 corresponds to majority holes, band
2 to majority electrons and band 3 to minority holes.
The insulating-like behavior of the carrier density with a tendency towards
saturation at low temperatures is well reproduced.
For the majority
bands, 1 and 2, the carrier concentrations are approximately equal and
similar in magnitude to literature values \cite{brandt}. The slope of
the linear-in-$T$ part of $\tau ^{-1}$=$\alpha _{\text{exp}}k_{B}T/\hbar $
with $\alpha _{\text{exp}}=0.065(3)$ (dashed line in Fig.~4, top panel)
is consistent with the electron-phonon mechanism of
scattering. To see this, we adopt a simple model in which carriers occupying
the ellipsoidal Fermi surface with parameters $m_{ab}$ (equal to 0.055$m_{0}$
and 0.040$m_{0}$ for electrons and majority holes, correspondingly) and $m_{c}$
(equal to $3m_{0}$ and $6m_{0},$ correspondingly) interact with longitudinal
phonons via a deformation potential, characterized by the coupling constant $%
D$ (equal to $27.9$ eV) . In this model, the slope in the linear-in-$T$
dependence of $\tau^{-1}$ is given by \cite{levinson} $\alpha _{\text{theor}}=\left(
\sqrt{2}/\pi \right) \sqrt{\left( m^{\ast }\right) ^{3}E_{F}}D^{2}/\rho
_{0}s_{ab}^{2}\hbar ^{3}$, where $m^{\ast }=\left( m_{ab}^{2}m_{c}\right)
^{1/3}\approx 0.21m_{0}$ both for electrons and holes, $\rho _{0}$=2.27 g/cm$%
^{3}$ is the mass density of graphite, and $s_{ab}=2\times 10^{6}$ cm/s is
the speed of sound in the ab-plane. (The numerical values of all parameters
are taken from the standard reference on graphite \cite{brandt}.) With the
above choice of parameters, $\alpha _{\text{theor}}=0.052$ for both types
of carriers. This value is within 20\% of the value found experimentally.
Given the simplicity of the model and the uncertainty in many material
parameters, especially in the value of $D$, such an agreement between
theory and experiment is quite satisfactory.

The solid lines through the data points in Fig.~1 are calculated
from the temperature-dependent fitting parameters derived from our
three-band analysis and plotted in Fig.~4. The shaded region (II)
depicted in the inset of Fig.~1 represents those temperatures and
fields that satisfy the inequalities in (1). In region (I) SdH
oscillations can be seen at sufficiently low $T$. Our sample has a
Dingle temperature of 3.0~K. The boundary between (I) and (II)
reflects the rightmost inequality of (1) and is determined by the
relation $T > \hbar eB/ m^{\ast}$.
The boundary between (II) and (III)
reflects the leftmost inequality of (1) and is determined by the
relation $B > m^{\ast}/e \tau (T)$ where $1/ \tau (T)$ is obtained
from experimental fitting parameters (Fig.~4). In the main panel
of Fig.~1 we superimpose region (II), again as a shaded area, on
the $\rho_{xx} (T,B)$ data. Below the lower boundary $\omega_c
\tau < 1$, and the magnetoresistance is relatively small. The
upper boundary is determined by the locus of ($B,T$) points
satisfying the rightmost inequality of (1). Clearly
region (II), constrained by the inequalities of (1), overlaps well
with the metal-insulating like behavior of graphite. 
Since the majority bands of bismuth comprise three
non-coplanar electron ellipsoids and one hole
ellipsoid, a similar analysis for bismuth is more complicated and
would require more space than available here.

We thus
conclude that the semimetals graphite and, by implication, bismuth
share the common features of high purity, low carrier density,
small effective mass and near perfect compensation and accordingly
obey the unique energy scale constraints that allow pronounced
metal insulating behavior accompanied by anomalously high
magnetoresistance. At magnetic fields higher than discussed in this
paper ($B \geq 1$~T) we believe that the multiband model is still appropriate
and may provide an alternative explanation for the reentrant behavior observed
by us and others \cite{kop3}.

Subsequent to the completion of this study, we were informed of
recent work [T. Tokumoto {\it et al.}, Solid State Commun. 129,
599 (2004)] which used a two-band model to explain the
unusual behavior of $\rho_{xx} (T,B)$ in graphite.

This work was supported in part by the National Science Foundation under
Grants No. DMR-0101856 (AFH) and DMR-0308377 (DLM) and by the In House Research
Program of the National High Magnetic Field Laboratory, which is suported by
NSF Cooperative Agreement No. DMR-0084173 and by the State of Florida.
SWT and DLM acknowledge the hospitality of the Aspen Center
for Theoretical Physics, where part of the analysis was performed.
The authors are especially thankful to J. E. Fischer and R. G. Goodrich
for supplying respectively the HOPG and Bi samples, J. Derakhshan for assistance
with the Bi measurements, and J. S. Brooks for discussions.


\begin{thebibliography}{99}
\bibitem[*]{tsai} Department of Physics, Boston University,
590 Commonwealth Ave., Boston, MA 02215.
\bibitem{edelman}  V. S. Edel'mann, Sov. Phys. Usp. 20, 819 (1977).

\bibitem{brandt}  N. B. Brandt, S. M. Chudinov, and Y. G. Ponomarev,
Semimetals 1: Graphite and its compounds (North-Holland, 1988).

\bibitem{kop1}  Y. Kopelevich et al., Phys. Solid State 41, 1959 (1999).

\bibitem{kop2}  H. Kempa, et al., Solid State Commun. 115, 539 (2000),
ibid., 121, 579 (2002).

\bibitem{kop3}  Y. Kopelevich et al., Phys. Rev. Lett. 90, 156402\/1 (2003).

\bibitem{khv}  D. V. Khveshchenko, Phys. Rev. Lett. 87, 206401 (2001).

\bibitem{kravchenko}  E. Abrahams, S. V. Kravchenko, and M. P. Sarachik,
Rev. Mod. Phys. 73, 251 (2001).

\bibitem{ashcroft}  N. W. Ashcroft and N. D. Mermin, Solid Sate Physics
(Holt, Rinehart and Winston, 1976).

\bibitem{abrikosov}  A. A. Abrikosov, Fundamentals of the Theory of Metals
(North-Holland, 1988).

\bibitem{levinson}  V. F. Gantmakher and Y. B. Levinson, Carrier scattering
in metals and semiconductors (North-Holland, 1987).

\bibitem{comment}  Inequality Eq.(\ref{a1}) can be satisfied in a typical metal
for $T\ll\Theta^\rho_D$
when the inverse (transport) time $\tau^{-1}_{\rm tr}\propto T^5\ll k_BT/\hbar$.
For an uncompensated metal with a closed Fermi surface, however,
magnetoresistance saturates in this regime.

\bibitem{mcclure}  J. W. McClure and W. J. Spry, Phys. Rev. 165, 809 (1968).

\bibitem{smith}   G. E. Smith, G. A. Baraff, and J. M. Rowell, Phys. Rev.
135, A1118 (1964).

\bibitem{cho}   S. Cho, Y. Kim, A. J. Freeman, et al., App. Phys. Lett.
79, 3651 (2001).

\bibitem{yang}   F. Y. Yang, K. Liu, K. Hong, et al., Science 284, 1335
(1999).

%
%
%
%
%

\end{thebibliography}
\end{document}